\documentclass{revtex4}
\usepackage{graphicx}
\begin{document}

\title{ A Five Dimensional Perspective on Many Particles \\ in  the Snyder basis of Double Special Relativity }
\author{J. M. Lorenzi$^{1}$, R. Montemayor$^{1}$  and L. F. Urrutia $^{2,3}$}
\affiliation{$^{1}$ Instituto Balseiro and CAB, Universidad Nacional de Cuyo and CNEA,
8400 Bariloche, Argentina}
\affiliation{$^{2}$ Instituto de Ciencias Nucleares, Universidad Nacional Aut{\'o}noma de
M{\'e}xico, A. Postal 70-543, 04510 M{\'e}xico D.F. \\}
\affiliation{$^{3}$ Facultad de F\'\i sica, Pontificia  Universidad Cat\'olica de Chile, Casilla 306, Santiago, Chile}

\begin{abstract}
After a brief summary of Double Special Relativity (DSR), we concentrate on a five dimensional procedure, which consistently introduce coordinates and momenta in the corresponding four-dimensional phase space, via a Hamiltonian approach. For the one particle case, the starting point is a de Sitter momentum space in five dimensions, with an additional constraint selected to recover the mass shell condition in four dimensions. Different basis of DSR can be recovered by selecting specific gauges to define the reduced four dimensional degrees of freedom. This is shown for the Snyder basis in the one particle case. We generalize the method to the many particles case and apply it again to this basis. We show that the energy and momentum of the system, given by the dynamical variables that are generators of translations in space and time and which close the Poincaré algebra, are additive magnitudes. From this it results that the rest energy (mass) of a composite object does not have an upper limit, as opposed to a single component particle which does.
 
\end{abstract}

\maketitle

\section{Introduction}
Double Special Relativity (DSR), or Special Relativity with two invariants,
arose originally \cite{GAC0} as an attempt to describe modified dispersion relations of
particles, presumably originating as a low energy consequence of quantum
gravity modifications of space-time, in a way consistent with a relativity
principle; i. e. without the need of introducing a preferred coordinate system or
ether \cite{GAC}.

One of the simplest examples is provided by the model of Magueijo and Smolin
which is  characterized by  the modified dispersion relation \cite{MASM}
\begin{equation}
\frac{1}{(1-\kappa p_{0})^{2}}\eta ^{\mu \nu }p_{\mu }p_{\nu }=m^{2}c^{2},
\label{DRMS}
\end{equation}%
that reduces to the standard Lorentz case when $\kappa \rightarrow 0.$ The
model is constructed by deforming the Lorentz group algebra in momentum
space, with generators
\begin{equation}
L_{\mu \nu }=p_{\mu }\frac{\partial }{\partial p^{\nu }}-p_{\nu }\frac{%
\partial }{\partial p^{\mu }},
\end{equation}
in such a away that the rotation generators $J^{i}=\epsilon ^{ijk}L_{jk}\;$%
are kept the same, while the boost generators are changed to
\begin{equation}
K^{i}=L_{0}^{\;\;i}+\kappa p^{i}p^{\mu }\frac{\partial }{\partial p^{\mu }}%
=U^{-1}(p_{0})L_{0}^{\;\;i}U(p_{0}),\qquad U(p_{0})p_{\mu }=\frac{1}{%
(1-\kappa p_{0})}p_{\mu }.\;\;
\end{equation}
The above modification preserves the standard Lorentz algebra among $%
J^{i},K^{j}$ but the Lorentz transformations are now realized non-linearly
in the form%
\begin{equation}
W[\omega ^{\mu \nu }]=U^{-1}(p_{0})\exp (\omega ^{\mu \nu }L_{\mu \nu
})U(p_{0}).
\end{equation}%
When applied to the momentum, the transformation for a boost in the
z-direction produces
\begin{eqnarray}
&& p_{0}^{\prime }=\frac{1}{D}\gamma (p_{0}-vp_{z}),\qquad p_{z}^{\prime }=%
\frac{1}{D}\gamma (p_{z}-vp_{0}),\qquad p_{x}^{\prime }=\frac{p_{x}}{D},
\qquad p_{y}^{\prime }=\frac{p_{y}}{D},  \label{BOOST} \\
&& D=1+\kappa (\gamma -1)p_{0}-\kappa \gamma vp_{z},
\end{eqnarray}%
which indeed preserve the dispersion relation (\ref{DRMS}). The main point
of the transformation (\ref{BOOST}) is that preserves the energy $%
E_{0}=1/\kappa$ under the corresponding transformations \cite{MASM}.
Also this energy is a maximum energy, as can be \ seen in the case of a
massive elementary particle. The energy and the momentum of such particle in
an arbitrary frame with velocity $\mathbf{v}$ are given by%
\begin{equation}
E=\frac{m_{0}\gamma }{1+\frac{m_{0}\gamma }{E_{0}}},\qquad \mathbf{p}=\frac{%
m_{0}\gamma }{1+\frac{m_{0}\gamma }{E_{0}}}\mathbf{v}
\end{equation}%
and we can verify that $E(m_{0}\gamma )\leq E_{0}$. Usually $E_{0}$ is taken
as the Planck energy $E_{P}$. At first sight this produces a contradiction
with the existence of composite macroscopic particles , which can certainly
have energies much larger than $E_{P}$. This is the so called '' soccer ball
problem'' and its resolution has to do with the fact that the energy in this model  is
not additive, due to the appearance of the non-linear transformations.

The idea of deforming the Poincar\'e algebra to produce alternative modified
dispersion relations has been generalized under the assumption that the only
accepted modification arises in the boost sector and that it is compatible
with rotation invariance. Under these conditions, the most general
modification is%
\begin{equation}
\left[ K_{i},p_{0}\right] =Cp_{i},\;\;\;\left[ K_{i},p_{j}\right] =A\delta
_{ij}+Bp_{i}p_{j}\;+D\epsilon _{ijk}p_{k},\;\;  \label{DEFORMATIONS}
\end{equation}%
where the functions $A,B,C,D\;$depend only on $p_{0}$, $\mathbf{p}^{2}\;$%
(scalars under rotation) and on the parameter $\kappa ,\;$in such a way
that the standard Poincar\'e limit is recovered when $\kappa \rightarrow 0$.
\ For example, the choice%
\begin{equation}
C=i,\;\;\;\;D=0,\;\;\;\;B=-i\kappa ,\;\;A=i\left( \frac{1}{2\kappa }%
(1-e^{-2p_{0}\kappa })+\frac{\kappa }{2}\mathbf{p}^{2}\right) ,
\end{equation}%
leads to the invariant dispersion relation%
\begin{equation}
2\frac{\cosh (\kappa p_{0})-1}{\kappa ^{2}}-\mathbf{p}^{2}e^{\kappa
p_{0}}=m^{2},
\end{equation}%
together with explicit expressions for the modified transformations in
momentum space. The above deformation defines what is called the
bicrossproduct basis in the literature \cite{BICROSS}.

One of the main drawbacks of this approach is the lack of information
regarding the coordinate space, which leads to ambiguities in the definition
of the velocity of the particle. Also, one would like to have a classical
version of the theory, in terms of a phase space endowed with a symplectic
structure. This will be the subject of Sections 3 and 4.

\section{ The five dimensional point of view}
One of the challenges faced in DSR is the construction of an appropriate
phase endowed with  coordinates and momenta that allow for the
description of events in inertial frames and  which transformations laws include
an additional universal invariant length (or energy) scale, besides the
standard invariant light velocity. Leaning on the analogy that the
formulation of Lorentz invariance in three dimensions, appropriate to the
description of a point particle having three degrees of freedom, gets
drastically simplified when going to four dimensions, a main road suggested
in the case of DSR is to start from a curved five dimensional space. In the
standard Lorentz case only one first class constraint is required to recover
the three degrees of freedom, while in this case we will need two first class
constraints to do the job.

The simplest case is to start with a de Sitter space, which is defined as a
four dimensional surface embedded in a five dimensional flat momentum space
according to \cite{SNYDER, KG1, FGEL,LORENZI}
\begin{equation}
\eta _{MN}P^{M}P^{N}+\kappa ^{2}=0.  \label{CURVEDME}
\end{equation}%
Here $M,N=0,1,2,3,4\;$and $\eta _{MN}=diag(1,-1,-1,-1,-1)$. The set of
transformations that leave the above surface invariant is the de Sitter
group $SO(4,1)$, which algebra is given by
\begin{equation}
\left[ l_{MN},l_{PQ}\right] =\eta _{MP}l_{QN}+\eta _{MQ}l_{NP}+\eta
_{NP}l_{MQ}+\eta _{NQ}l_{PM}.  \label{GENALG}
\end{equation}%
In this way, $\kappa$ is interpreted as the invariant energy. The generators $%
l_{PQ}$ have the following matrix realization%
\begin{equation}
\left[ l_{PQ}\right] _{\;\;N}^{M}=\delta _{P}^{M}\eta _{QN}-\delta
_{Q}^{M}\eta _{PN}.
\end{equation}%
The generators act in momentum space as%
\begin{equation}
P^{M}\rightarrow P^{\prime M}=\left[ \exp (\theta _{PQ}l_{PQ})\right]
_{\;\;N}^{M}P^{N}.
\end{equation}%
Using the relation
\begin{equation}
\left\{ P^{M},L_{PQ}\right\} =\left( \frac{\partial P^{\prime M}}{\partial
\theta _{PQ}}\right) _{\theta =0},
\end{equation}%
to define the group action in momentum space through the bracket\ $\left\{
{}\right\} $, one readily obtains
\begin{equation}
\left\{ P^{M},L_{NQ}\right\} =\delta _{N}^{M}P_{Q}-\delta _{Q}^{M}P_{N}.
\end{equation}%
Also, it can be shown that the brackets $\left\{ L_{MN},L_{PQ}\right\} \;$%
inherit the algebra (\ref{GENALG}). In the following it proves convenient to
make the splitting $\;\left\{ M\right\} =\left\{ \mu ,4;\;\mu
=0,1,2,3\right\} $. The notation is
\begin{equation}
L_{k}=\frac{1}{2}\epsilon _{klm}L_{lm},\;\;\;B_{i}=L_{0i},\;\;\;D_{\mu
}=L_{\mu 4}.
\end{equation}
One consequence of this five-dimensional starting point is the appearance of
five additional symmetries ($P_{4},D_{\mu }$), besides the usual ten
symmetries ($3$ rotations + $3$ boosts + $4$ translations) which describe
the transformations among inertial frames. A possible interpretation of such
extra symmetries will be given in Section 4.

Starting from this five dimensional perspective there are at least two roads
to construct the required phase space: (1) to search for a realization of
the physical coordinates $x^{\mu }$  as combinations of appropriate
generators of the de Sitter group  in momentum space and (2) to
build DSR as a constrained theory in this five-dimensional space. We
present a brief review of the first method in this section and develop the
second in the following sections, including the one and many particles
cases.

Approach (1) is rooted in the work of  Snyder \cite{SNYDER}, who was the first in
introducing   a fundamental invariant length by starting from the curved
momentum space (\ref{CURVEDME}). Later, Kowalski-Glikman showed that the
Snyder construction can be generalized to include all deformations described
in Eq. (\ref{DEFORMATIONS}) \cite{KG1}. The basic idea is to define a map
\begin{equation}
x^{\mu }=x^{\mu }\left( P_{M},\;L_{PQ},\kappa \right) ,\;\;\;\;p_{\nu
}=p_{\nu }(P_{M},\kappa ),
\end{equation}%
which recovers the phase space version of a given deformed algebra together
with the corresponding invariant dispersion relation. Two typical examples
of this procedure are the construction of (i) the Snyder basis and (ii) the
bicrossproduct basis, which correspond to the choices
\begin{equation}
x_{\mu }=-\frac{D_{\mu }}{\kappa },\;\;p_{\mu }=\kappa \frac{P_{\mu }}{%
P_{4}},
\end{equation}%
and
\begin{equation}
x_{0}=-\frac{D_{0}}{\kappa },\;\;\;\;x_{i}=-\frac{1}{\kappa }%
(B_{i}+D_{i}),\;\;\;p_{0}=\kappa \ln \left( \frac{P_{4}-P_{0}}{\kappa }%
\right) ,\;\;\;\;\;p_{i}=\frac{\kappa P_{i}}{P_{0}-P_{4}},
\end{equation}%
respectively.

As we can see, this method provides no criteria to single out some specific
choice which could be subsequently subjected to experimental/observational
verification. Also, it does not provide any natural definition of the
velocity.
\section{ The one-particle case}
As we showed in the previous sections, it is possible to introduce an invariant length (or energy) in the
transformations that connect inertial frames  by starting from a momentum space with constant curvature.
Nevertheless, the definition of the corresponding phase space, together with that of the particle velocity
remains an open problem due to the many possibilities that arise. In this and the following sections we discuss
a method which provides a unified version of the many alternatives already present. In complete analogy with the formulation of the relativistic particle in four dimensions, the basic idea is to view a DSR
particle as arising from a constrained system in five dimensions, defined through a first order action which includes the
concepts of coordinates and velocities from the very beginning. Since the initial phase space has now  ten degrees of freedom, we will require two first class constraints (as opposed to one first class constraint in the relativistic case) in order to recover the final three degrees of freedom in coordinate space.  As we will explain in the following, the different basis
previously discussed, and many others, arise in this formulation as the result of different  gauge fixings for  one of the first class constraints. This method, applied to the one-particle case, has been previously discussed in Ref. \cite{FGKG}. 

Our starting point is the five-dimensional action
\begin{equation}
S=\int d\tau \left({\dot X}^M\eta_{MN}P^N - \Lambda H_{5d}- \lambda H_{4d} \right),
\end{equation}
where $\Lambda$ and $\lambda$  are Lagrange multipliers. Here $\tau$ is the proper time and ${\dot A}=dA/d\tau$. The constraints are
\begin{equation}
H_{5d}=P^MP_M+ \kappa^2, \qquad H_{4d}=P^\mu P_\mu -m^2,
\end{equation}
where we have chosen $H_{4d}$ as  the four-dimensional mass shell condition for a particle with mass $m$. This constraint can be more conveniently written as
\begin{equation}
H_{4d}=P^4-M,\qquad M=\sqrt{m^2+\kappa^2},  \qquad P^4=\sqrt{P^\mu P_\mu + \kappa^2}.
\end{equation}
Introducing a small change in notation ($X_4=z_4, \quad P^4=\xi^4$) the initial action is now written as
\begin{equation}
S=\int d\tau \left({\dot z}_4 \xi^4+ {\dot X}_\mu P^\mu  - \Lambda \left( \xi_4 \xi^4+ P_\mu P^\mu +\kappa^2  \right)
- \lambda \left( \xi^4-\sqrt{m^2+\kappa^2}\right)\right).
\end{equation}
We start with 12 coordinates: $z_4, X_\mu, \xi^4, P^\mu, \Lambda, \lambda$; together with their respective momenta ${\Pi}_z^4, {\Pi}_X^\mu, {\Pi}_\xi^4,  {\Pi}_P^\mu, {\Pi}_\Lambda, {\Pi}_\lambda$; satisfying the standard Poisson brackets for canonical variables. The primary constraints are
\begin{equation}
{\Pi}_z^4-\xi^4 \approx 0, \quad {\Pi}_X^\mu-P^\mu\approx 0, \quad  {\Pi}_\xi^4\approx 0, \quad {\Pi}_P^\mu\approx 0, \quad
{\Pi}_\Lambda\approx 0,\quad {\Pi}_\lambda\approx 0.
\end{equation}
The extended Hamiltonian is
\begin{eqnarray}
H=\Lambda\left(  \xi_{4}\xi^{4}+P_{\mu}P^{\mu}+\kappa^{2}\right)
+\lambda\left(  \xi^{4}-\sqrt{m^{2}+\kappa^{2}}\right)
 +a_{4}\left(  \Pi_{z}^{4}-\xi^{4}\right)  +b_{\mu}\left(  \Pi_{X}^{\mu
}-P^{\mu}\right)  +d_{4}\Pi_{\xi}^{4}+e_{\mu}\Pi_{P}^{\mu}+g\Pi_{\Lambda}%
+h\Pi_{\lambda}.
\end{eqnarray}
where $a_4, b_\mu, d_4, e_\mu , g, h$ are arbitrary functions. The conservation of the primary constraints fixes some arbitrary functions
\begin{eqnarray}
&&\left\{  \Pi_{z}^{4}-\xi^{4},H\right\}\approx 0 \quad \rightarrow \quad  d_{4}=0, \qquad
\left\{  \Pi_{X}^{\mu}-P^{\mu},H\right\} \approx 0 \quad \rightarrow \quad e_{\mu}=0, \\
&&\left\{  \Pi_{\xi}^{4},H\right\} \approx 0,  \quad
\rightarrow \quad a^{4}=2\Lambda\xi^{4}-\lambda,  \quad
\left\{  \Pi_{P}^{\mu},H\right\}\approx 0 \quad \rightarrow \quad
b^{\mu}=2\Lambda P^{\mu}.
\end{eqnarray}
and also provides the following secondary constraints
\begin{eqnarray}
\left\{  \Pi_{\Lambda},H\right\}=\xi_{4}\xi^{4}+P_{\mu}P^{\mu}%
+\kappa^{2}\approx 0, \qquad
\left\{  \Pi_{\lambda},H\right\}=\xi^{4}-\sqrt{m^{2}+\kappa^{2}}\approx 0.
\end{eqnarray}
The  secondary  constraints turn out to be automatically conserved. In this way the Hamiltonian results
\begin{eqnarray}
H =\Lambda\left(  \xi_{4}\xi^{4}+P_{\mu}P^{\mu}+\kappa^{2}\right)
+\lambda\left(  \xi^{4}-\sqrt{m^{2}+\kappa^{2}}\right)
 +\left(  2\Lambda\xi-\lambda\right)  \left(  \Pi_{z}^{4}-\xi^{4}\right)
+2\Lambda P_{\mu}\left(  \Pi_{x}^{\mu}-P^{\mu}\right)  +g\Pi_{\Lambda}%
+h\Pi_{\lambda}.
\end{eqnarray}
The above constraints can be further classified into four  first class constraints
\begin{eqnarray}
&\Pi_{\Lambda}\approx 0, \qquad
\Pi_{\lambda} \approx 0, \\
&2\left(  \xi_{4}\Pi_{z}^{4}+p_{\mu}\Pi_{x}^{\mu}\right)  -\left(  \xi
_{4}\xi^{4}+p_{\mu}p^{\mu}\right)  +\kappa^{2}\approx 0, \\
&\Pi_{z}^{4}-\sqrt
{m^{2}+\kappa^{2}}\approx 0
\end{eqnarray}%
and ten second class constraints
\begin{eqnarray}
\Pi_{z}^{4}-\xi^{4}   \approx 0, \qquad
\Pi_{X}^{\mu}-P^{\mu}   \approx 0, \qquad
\Pi_{\xi}^{4}   \approx 0, \qquad
\Pi_{P}^{\mu} \approx 0.
\end{eqnarray}
We can now verify the count of degrees of freedom (DOF) in coordinate space, which is
\begin{equation}
\# \, DOF=\frac{1}{2}\left( 2\times 12-2\times 4-10  \right)=3,
\end{equation}
as expected.
After imposing  strongly the second class constraints, we are left with the Hamiltonian
\begin{equation}
H =\Lambda\left( \xi_{4}\xi^{4}+P_{\mu}P^{\mu}+\kappa^{2}\right)
+\lambda\left(  \xi^{4}-\sqrt{m^{2}+\kappa^{2}}\right)
+g\Pi_{\Lambda}+h\Pi_{\lambda},
\end{equation}
together with the  first class constraints
\begin{eqnarray}
\Pi_{\Lambda} \approx 0, \qquad
\Pi_{\lambda} \approx 0, \qquad
H_{5d}=\xi_{4}\xi^{4}+P_{\mu}P^{\mu}+\kappa^{2} \approx 0, \qquad
H_{4d}=\xi^{4}-\sqrt{m^{2}+\kappa^{2}} \approx 0.
\end{eqnarray}
The   Dirac brackets among the remaining phase space variables turn out to be  identical with the original
Poisson brackets. In an abuse of notation, we do not label with additional indexes the resulting Dirac brackets that appear in each step of the calculation. The non-zero values are
\begin{equation}
\left\{z_4, \xi^4\right\}=1, \qquad \left\{X_\mu, P^\nu\right\}=\delta^{\nu}_{\mu}, \qquad \left\{\Lambda, \Pi_\Lambda\right\}=1,
\qquad \left\{\lambda, \Pi_\lambda\right\}=1.
\end{equation}
Before dealing with  particular cases, we state some general requirements to implement such procedure. Since we have four first class constraints we will require to add four additional constraints $\chi^1, \chi^2, {\bar {\chi}}^1, {\bar{\chi}}^2$, which must have  zero Poisson bracket with the Hamiltonian,  and be such that the whole set of eight constraints is now second class.

We will make use of the iterative method of sequentially fixing the gauge, calculating in each step the resulting  Dirac brackets. First we eliminate the variables $\Lambda, \lambda, \Pi_\Lambda, \Pi_\lambda$. To this end we take
\begin{equation}
{\bar{\chi}}^1=\Lambda-\bar{\Lambda}\left(z_4, X_\mu, \xi^4, P^\mu \right)\approx 0,  \qquad
{\bar{\chi}}^2=\lambda-\bar{\lambda}\left(z_4, X_\mu, \xi^4, P^\mu \right)\approx 0.
\end{equation}
The time evolution $d{\bar{\chi}}^{1,2}/d\tau=0$ fixes the arbitrary functions  $g$ and $h$. The remaining canonical variables are $z_4, X_\mu, \xi^4, P^\mu$ with non zero Dirac brackets
\begin{equation}
\left\{z_4, \xi^4\right\}=1, \qquad \left\{X_\mu, P^\nu\right\}=\delta^{\nu}_{\mu}
\end{equation}
and Hamiltonian
\begin{equation}
H =\bar{\Lambda}\left( \xi_{4}\xi^{4}+P_{\mu}P^{\mu}+\kappa^{2}\right)
+\bar{\lambda}\left(  \xi^{4}-\sqrt{m^{2}+\kappa^{2}}\right),
\end{equation}
together with the constraints
\begin{eqnarray}
H_{5d}=\xi_{4}\xi^{4}+P_{\mu}P^{\mu}+\kappa^{2} \approx 0, \qquad
H_{4d}=\xi^{4}-\sqrt{m^{2}+\kappa^{2}} \approx 0, \qquad \chi^1 \approx 0, \qquad \chi^2 \approx 0.
\end{eqnarray}
As we will show in the following, some of the previously discussed  DSR basis can be obtained by choosing different gauge  fixings
for the constraint $H_{5d}$.
The strategy is the following. Since we want to impose strongly $H_{5d}$ and $\chi^1$ we require that
\begin{equation}
\{H_{5d}, \chi^1 \}=C(X,P)\neq 0,
\label{GAUGEFIX}
\end{equation}
with
\begin{equation}
\{H, \chi^1 \}= \bar{\Lambda}\,C+ \bar{\lambda}\, \{H_{4d}, \chi^1 \} \approx 0,
\label{CONSIST}
\end{equation}
 which determines the relation between $\bar{\Lambda}$ and $\bar{\lambda}$. Taking into account the constraints $H_{5d}$ and $\chi^1$ we next define an invertible coordinate transformation
\begin{equation}
\left( \xi_4, P_\mu, z^4, X^\mu \right) \rightarrow \left( p_\mu, H_{5d}, x^\mu, \chi^1 \right),
\end{equation}
where
\begin{equation}
x^\mu=x^\mu(\xi_4, P_\mu, z^4, X^\mu), \qquad p_\mu=p_\mu(\xi_4, P_\mu, z^4, X^\mu),
\label{COORDT}
\end{equation}
which are functions of the five-dimensional phase space, are what we define as  the physical four dimensional phase space coordinates.
We restrict ourselves to the case when
\begin{equation}
\left\{H_{5d}, x^\mu \right\}= \left\{ H_{5d}, p_\mu \right\}=\left\{\chi^1, x^\mu \right\}=\left\{\chi^1, p_\mu \right\}=0,
\label{GINVCOND}
\end{equation}
so that $x^\mu$ and $p_\mu$ are invariant under the gauge transformations generated by $H_{5d}$ and $\chi^1$. In this way, they become observable, whose dynamics will be determined by the remaining constraint $H_{4d}$ in the next step of the procedure. A first consequence of (\ref{GINVCOND}) is that the resulting Dirac  bracket in this step,
\begin{equation}
\left\{A, B\right\}_{DB}=\left\{A, B\right\}-\left\{A, \chi^1 \right\}\frac{1}{C}\left\{H_{5d}, B\right\}+
\left\{A, H_{5d} \right\}\frac{1}{C}\left\{\chi^1, B\right\},
\end{equation}
is equal to the previous bracket in the case of the dynamical four dimensional variables $x^\mu$ and $p_\mu$.	 Moreover, given any function $A(\xi_4, P_\mu, z^4, X^\mu)$ having  zero Poisson brackets with $H_{5d}$ and $\chi^1$, it will depend only upon $x^\mu$ and $p_\mu$ after making the inverse transformation (\ref{COORDT}). This can be seen by calculating
\begin{eqnarray}
&&0=\left\{A, H_{5d}\right\}=\frac{\partial A}{\partial x^\mu} \left\{x^\mu, H_{5d}\right\}+
\frac{\partial A}{\partial p_\mu} \left\{p_\mu, H_{5d}\right\}+
\frac{\partial A}{\partial \chi^1} \left\{\chi^1, H_{5d}\right\}=\frac{\partial A}{\partial \chi^1} \left\{\chi^1, H_{5d}\right\},\\
&& 0=\left\{A, \chi^1\right\}=\frac{\partial A}{\partial x^\mu} \left\{x^\mu, \chi^1\right\}+
\frac{\partial A}{\partial p_\mu} \left\{p_\mu, \chi^1\right\}+
\frac{\partial A}{\partial H_{5d}} \left\{H_{5d}, \chi^1 \right\}=-\frac{\partial A}{\partial H_{5d}} \left\{\chi^1, H_{5d}\right\}.
\end{eqnarray}
Since we require (\ref{GAUGEFIX}), the conclusion is that ${\partial A}/{\partial H_{5d}}$ and ${\partial A}/{\partial \chi^1}$ are zero. In this way we confirm that the variables $x^\mu$ and $p_\mu$ allow to completely specify any observable function in the constraint surface determined by $H_{5d}$ and $\chi^1$, which is where the dynamics occurs.
\subsection{ The Snyder coordinates}
In order to recover this basis we choose
\begin{equation}
\chi^1_S=X^\mu P_\mu +z^4 \xi_4-T.
\end{equation}
We can directly verify that $C=2
\kappa^2 \neq 0$ and that the condition (\ref{CONSIST}) reduces to $2 \bar{\Lambda} \kappa^2= \bar{\lambda} \xi_4 $.
The Snyder coordinates are subsequently defined as
\begin{equation}
p_\mu=\kappa\frac{P_\mu}{\xi_4}, \qquad  x_\mu=\frac{1}{\kappa}\left(X_\mu \xi_4 -z_4 P_\mu \right),
\label{DEFSNYDER}
\end{equation}
which satisfy the condition (\ref{GINVCOND}). The inversion of the above definition produces
\begin{equation}
P_\mu=\frac{p_\mu}{\Theta}, \quad \xi_4=\frac{\kappa}{\Theta} , \quad z_4=\frac{x^\mu p_\mu-T}{\kappa \Theta},
\quad X^\mu=\Theta x^\mu + \frac{p^\mu}{\kappa \Theta}\left( x^\alpha p_\alpha -T \right), \quad \Theta= \sqrt{1-p^\alpha P_\alpha/\kappa^2}.
\end{equation}
The bracket algebra for the Snyder coordinates is calculated from the already established five-dimensional algebra, via the Eqs.(\ref{DEFSNYDER}), and yields
\begin{equation}
\left\{ x_\mu, x_\nu \right\}=-\frac{1}{\kappa^2}\left( x_\mu p_\nu-x_\nu p_\mu\right), \qquad
\left\{ x_\mu, p_\nu \right\}=\eta_{\mu\nu}-\frac{p_\mu p_\nu}{\kappa^2},  \qquad \left\{ p_\mu, p_\nu \right\}=0.
\label{BRACKETSC}
\end{equation}
The Hamiltonian is
\begin{equation}
H=\bar{\lambda} \left( \frac{C}{\sqrt{\kappa^2-p^\alpha p_\alpha}} -M \right)=\bar{\lambda} H_{4d}.
\end{equation}
The constraint $H_{4d}$ reduces to
\begin{equation}
H_{4d}=p^\alpha p_\alpha-\kappa^2\left( 1-\frac{\kappa^2}{M^2} \right),
\end{equation}
which is interpreted as the mass shell condition for a relativistic particle with mass $m$
\begin{equation}
m=\kappa \sqrt{1-\frac{\kappa^2}{M^2}}.
\end{equation}
Let us observe  that the above relation leads to $m=\kappa$ as the maximum mass accessible to such elementary particle. The equation of motions are
\begin{equation}
\frac{d p_\mu}{d\tau}=\left\{p_\mu, H \right\}=0,  \qquad \frac{d x_\mu}{d\tau}=\left\{x_\mu, H \right\}=\frac{{\bar \lambda} p_\mu}{\sqrt{\kappa^2-p^\alpha p_\alpha}}.
\end{equation}
leading to the velocity ($c=1$)
\begin{equation}
v^i=\frac{d x^i}{d t}=\frac{d x^i}{d \tau} \frac{d \tau}{d x^0}=\frac{p^i}{p^0}.
\end{equation}

\section{ The multiparticle case in Snyder coordinates}
In this section we proceed in analogy with the one particle case in order to build a consistent DSR
formulation for the case of $N$ particles. In the previous case we enlarged the original phase space to five dimensions
and imposed the constraint that the momentum sector has constant curvature $\kappa$, corresponding to a de Sitter space. Also we required that the particle satisfied a four
dimensional mass shell condition designed  to recover the three DOF in coordinate space. In this way we have introduced two invariant constants: the curvature $\kappa$ and the light velocity $c$.

The natural generalization for the  $N$ particles case is to start from a configuration space
$\left( X_{\mu}^{a}, z_{4}\right), \,  a=1,2,..,N,$ with dimension
$4N+1$. The additional coordinate  $z_4$ is taken to be spacelike. Here, each subset $\left( X_{\mu}^{a} \right)$
describes the position of particle $a$ having a mass $m_a$. The momentum space is labeled by $\left(P^\mu_a, \xi^4 \right)$
and the mass shell condition
$P_{a}^{\mu}{P_{a}}_{\mu}=m_{a}^{2}$ defines the universal speed of light $c=1$. There is an additional condition  that constrains the momentum space to an hypersurface with constant curvature $\kappa$.

Let us consider  the first order Lagrangian
\begin{equation}
L=\dot{z}_{4}\xi^{4}+\sum_{a=1}^{N}\dot{X}_{\mu}^{a}P_{a}^{\mu}-\Lambda\left(
\xi_{4}\xi^{4}+\sum_{a}P_{\mu}^{a}P_{a}^{\mu}+\kappa^{2}\right)  -\sum
_{a}\lambda_{a}\left(  P_{\mu}^{a}P_{a}^{\mu}-m_{a}^{2}\right),
\end{equation}
with
$P_{\mu}P^{\mu}=\left(  P^{0}\right)^{2}-P^{i}P^{i}$ for each particle.
From the de Sitter constraint and the dispersion relations we can write
\begin{eqnarray}
\xi^{4}=\sqrt{\kappa^{2} + \sum_{a}m_{a}^{2}} \equiv M >0.
\label{CONST1}
\end{eqnarray}
As a simplification, we replace the constraint for the first particle ($a=1$) by (\ref{CONST1}), in such a way that
we start from
\begin{eqnarray}
L =\dot{z}_{4}\xi^{4}+\sum_{a}\dot{X}_{\mu}^{a}P_{a}^{\mu}-\Lambda\left(
\xi_{4}\xi^{4}+\sum_{a=1}^N P_{\mu}^{a}P_{a}^{\mu}+\kappa^{2}\right)
-\lambda\left(  \xi^{4}-M\right)  -\sum_{b=2}^N\lambda_{b}\left(  P_{\mu}^{b}%
P_{b}^{\mu}-m_{b}^{2}\right).
\end{eqnarray}
The $9N+3$ coordinates are $z_4, \xi^4, x_\mu^a, p^\mu_a, \lambda, \lambda_b$.
The calculation of the canonical momenta produces the following primary constraints
\begin{eqnarray}
&& \Pi_{z}^{4} -\xi^{4} \approx 0, \quad
\Pi_{X_a}^{\mu}-P_{a}^{\mu} \approx 0,\quad
\Pi_{{\xi}^{4}}\approx 0, \quad
\Pi_{P_a}^{\mu}\approx 0, \quad
\Pi_{\Lambda}\approx 0, \quad
\Pi_{\lambda}\approx 0, \quad
\Pi_{{\lambda}_{b}}\approx 0.
\label{PRIMC}
\end{eqnarray}
The extended Hamiltonian is %
\begin{eqnarray}
&&H=\Lambda\left(\xi_{4}\xi^{4}+\sum_{a=1}^N P_{\mu}^{a}P_{a}^{\mu}+\kappa
^{2}\right)  +\lambda\left( \xi^{4}-M\right)  +\sum_{b=2}^N \lambda_{b}\left(
P_{\mu}^{b}P_{b}^{\mu}-m_{b}^{2}\right) \\
&&  + a_{4}\left(  \Pi_{z}^{4}-\xi^{4}\right)  + \sum_{a=1}^N b_{\mu}^{a}\left(
\Pi_{X_a}^{\mu}-P_{a}^{\mu}\right)  +d_{4}\Pi_{{\xi}^{4}}+\sum_{a=1}^N e_{\mu}^{a}%
\Pi_{P_a}^{\mu}+g\Pi_{\Lambda}+h\Pi_{\lambda}+\sum_{a=1}^Nj_{a}\Pi_{{\lambda}_{a}},
\end{eqnarray}
with $a_4, b_\mu^a, d_4, e_\mu^a, g, h, j_a$ being arbitrary functions. The conservation of the primary constraints
fixes some of the arbitrary functions
\begin{eqnarray}
&&\left\{\Pi_{z}^{4}-\xi^{4},H \right\}\approx 0 \,\, \rightarrow \,\, d_{4}=0, \qquad
\left\{\Pi_{{X_a}}^{\mu}-P_{a}^{\mu},H \right\}\approx 0 \,\, \rightarrow \,\,
e_{\mu}^{a}=0, \\
&& \left\{\Pi_{{\xi}^{4}},H \right\}  \approx 0 \,\, \rightarrow \,\,
a^{4}=2\Lambda\xi^{4}-\lambda, \qquad
\left\{  \Pi_{{P_{a}}}^{\mu},H\right\}\approx 0 \,\, \rightarrow \,\, b_{a}^{\mu}=2\Lambda
P_{a}^{\mu}+2\lambda_{a}P_{a}^{\mu}
\end{eqnarray}
and also generates the following secondary constraints
\begin{eqnarray}
&&\left\{ \Pi_{\Lambda},H\right\}\approx 0 \,\,  \rightarrow \,\, \xi_{4}\xi^{4}+\sum_{a=1}^N P_{\mu}^{a}%
P_{a}^{\mu}+\kappa^{2}\equiv H_{D+1}\approx 0, \nonumber \\
&&\left\{  \Pi_{\lambda},H\right\}\approx 0 \,\,  \rightarrow \,\,  =\xi^{4}-M\equiv H_{4}\approx 0, \qquad
\left\{\Pi_{\lambda}^{b},H\right\}\approx 0 \,\,  \rightarrow \,\,  P_{\mu}^{b}P_{b}^{\mu}-m_{b}^{2} \approx 0.
\label{SECONDC}
\end{eqnarray}
The Hamiltonian results
\begin{eqnarray}
&&H =\Lambda H_{D+1}+\lambda H_{4}+\sum_{b=2}^N \lambda_{b}\left(  P_{\mu}%
^{b}P_{b}^{\mu}-m_{b}^{2}\right) +\left(  2 \Lambda\xi^{4}-\lambda\right)  \left(  \Pi_{z}^{4}-\xi^{4}\right)\nonumber \\
&& +2\sum_{a=1}^N \Lambda {P_{a}}_{\mu}\left(  \Pi_{{X_a}}^{\mu}-P_{a}^{\mu
}\right)  +2\sum_{b=2}^N \lambda_{b}{P_{b}}_{\mu}\left(  \Pi_{b}^{\mu}-P_{b}^{\mu
}\right)  +g\Pi_{\Lambda}+h\Pi_{\lambda}+\sum_{a}j_{a}\Pi_{\lambda}^{a},
\end{eqnarray}
with the constraints given in (\ref{PRIMC}) together with ({\ref{SECONDC}).
The  first class constrains are
\begin{eqnarray}
&& \Pi_{\Lambda}\approx 0 , \qquad
\Pi_{\lambda_b} \approx 0, \qquad
\Pi_\lambda \approx 0, \qquad \Pi_{z}^{4}-M  \approx 0,
\nonumber \\
&& \frac{1}{2}H_{D+1}\equiv   \xi_{4}\pi_{z}^{4}+\sum_{a}P_{\mu}^{a}\Pi_{X_a}^{\mu}
+\kappa^{2}  \approx 0 ,\qquad
2{P_b}_{\mu}\Pi^{\mu}_{P_b}-{P_b}_{\mu}P_{b}^{\mu}-m_{b}^{2}  \approx 0, \quad  b=2,3,...,N,
\label{FCC}
\end{eqnarray}
while the remaining second class constraints are
\begin{eqnarray}
\Pi_{z}^{4}-\xi^{4}  \approx 0, \qquad
\Pi_{X_a}^{\mu}-P_{a}^{\mu}  \approx 0, \qquad
\Pi_{\xi}^{4}  \approx 0, \qquad
\Pi_{P_{a}}^{\mu} \approx 0.
\end{eqnarray}
We have  $9N+3$ coordinates with  $8N+2$ second class constraints and $2N+2$ first class constraints.
The count of the DOF is
\begin{equation}
\# DOF= \frac{1}{2}\left( 2(9N+3)-2(2N+2)-(8N+2) \right)=3N,
\end{equation}
which corresponds to the $N$ particles considered.

\subsection{ Imposing  the constraints strongly}
As in the previous section, we impose the constraints in a stepwise mode. Again, in an abuse of notation, in many occasions we will not introduce an additional notation for the modified Dirac brackets that arise at each step, unless some confusion arises.
First we start with the second class constraints
\begin{eqnarray}
\phi_{1} =\Pi_{z}^{4}-\xi^{4}, \qquad
\phi_{2}  &  =\Pi_{\xi^4}=0, \qquad
\phi_{3}^{\mu a} =\Pi_{X_a}^{\mu}-P_{a}^{\mu}, \qquad
\phi_{4}^{\mu a}=\Pi_{P_{a}}^{\mu},
\end{eqnarray}
which we use to eliminate the variables
\begin{equation}
\Pi_z^4,\, \Pi_{\xi^4},\,   \Pi_{X_a}^{\mu}, \, \Pi_{P_{a}}^{\mu}.
\label{ELIM1}
\end{equation}
The non-zero brackets are
\begin{eqnarray}
\left\{  \phi_{1},\phi_{2}\right\} =-1, \qquad
\left\{  \phi_{3}^{\mu a},\phi_{4}^{\nu c}\right\}=-\eta^{\mu\nu}%
\delta^{ac}, \qquad
\left\{  \phi_{5}^{\mu},\phi_{6}^{\nu}\right\}=-\eta^{\mu\nu}%
\end{eqnarray}
and one can verify that the (Dirac) brackets for the variables that  remain after having eliminated  those indicated in (\ref{ELIM1}) are equal to the initial Poisson brackets.
In the restricted hypersuface determined by the second class constraints we have the Hamiltonian
\begin{equation}
H=\Lambda H_{D+1}+\lambda H_{4}+\sum_{b=2}^N\lambda_{b}\left(  {P_{\mu}}^{b}%
{P^{\mu}}_{b}-m_{b}^{2}\right)  +g\Pi_{\Lambda}+h\Pi_{\lambda}+\sum_{a=1}^N j_{a}%
\Pi_{{\lambda}_a},
\end{equation}
plus the first class constraints in (\ref{FCC}). After  the imposition of the second class constraints the first class constraints now read
\begin{eqnarray}
&&\Pi_\Lambda \approx 0, \qquad \Pi_{\lambda_b}\approx 0, \qquad \Pi_\lambda\approx 0, \qquad H_4\equiv \xi^4-M \approx 0, \nonumber \\
&&\frac{1}{2}H_{D+1}\equiv \xi^4\xi_4 + \sum_{a=1}^N P^a_\mu P_a^\mu +\kappa^2\approx 0, \qquad {P_b}_{\mu}P_{b}^{\mu}+m_{b}^{2}  \approx 0,
\end{eqnarray}

\subsection{ Fixing the gauge in the first class constraints}
Let us denote  the gauge fixing conditions by
$\chi_{\Lambda}$,\, $\chi_{\lambda_b}$, \, $\chi_{\lambda}$, \, $\chi_{H_4}$, \, $\chi_{H_{D+1}}$,\,
 $\chi_{{P}_{b}}$ respectively. Taking into account the auxiliary character of the variables
$\Lambda$, $\lambda$, $\lambda_{b}$, we demand that the gauge fixing constraints  depend upon them according to the following scheme
\begin{eqnarray}
&&\left\{  \chi_{\Lambda},\Pi_{\lambda}\right\}=\left\{  \chi_{\Lambda
},\Pi_{\lambda_b}\right\}  =0, \qquad
\left\{  \chi_{\lambda},\Pi_{\Lambda}\right\} =\left\{  \chi_{\lambda
},\Pi_{\lambda_b}\right\}=0, \qquad
\left\{  \chi_{\lambda_b},\Pi_{\Lambda}\right\} =\left\{  \chi_{\lambda_b},\Pi_{\lambda
}\right\}  =0, \nonumber \\
&&\left\{  \chi_{_{H_{D+1}}},\Pi_{\lambda}\right\}=\left\{  \chi_{H_{D+1}}%
,\Pi_{\lambda_b}\right\}  =\left\{  \chi_{H_{D+1}},\Pi_{\Lambda}\right\}=0, \nonumber \\
&&\left\{  \chi_{\xi},\Pi_{\lambda}\right\}=\left\{  \chi_{\xi},\Pi
_{\lambda_b}\right\} =\left\{  \chi_{\xi},\Pi_{\Lambda}\right\}=0, \nonumber \\
&&\left\{ \chi_{{P}_{b}},\Pi_{\lambda}\right\}=\left\{ \chi_{P_{b}},\Pi_{\lambda_b}\right\}
 =\left\{  \chi_{{P}_{b}},\Pi_{\Lambda}\right\} =0.
\end{eqnarray}
Using the above prescription, in order to eliminate the auxiliary variables $\Lambda$, $\lambda$, $\lambda_{b}$
we choose
\begin{eqnarray}
\chi_{\Lambda} =\Lambda-\tilde{\Lambda}\left(  z^{4},\xi^{4},X_{a}^{\mu
},P_{a}^{\mu}\right), \qquad
\chi_{\lambda} =\lambda-\tilde{\lambda}\left(  z^{4},\xi^{4},X_{a}^{\mu
},P_{a}^{\mu}\right) \qquad
\chi_{\lambda_b}=\lambda_{b}-\tilde{\lambda}_{b}\left(  z^{4},\xi^{4},X_{a}^{\mu
},P_{a}^{\mu}\right).
\end{eqnarray}
Demanding conservation of the above constraints we determine the arbitrary functions $g, \, h, \, j_b $
in such a way that, in the new restricted space, the Hamiltonian is
\begin{equation}
H=\tilde{\Lambda}H_{D+1}+\tilde{\lambda}H_{4}+\sum_{b}\tilde{\lambda}%
_{b}\left(  p_{\mu}^{b}p_{b}^{\mu}-m_{b}^{2}\right).
\end{equation}
The set $\left(  \Pi_\Lambda,  \Pi_\lambda, \Pi_{\lambda_b}, \chi_\Lambda, \chi_\lambda, \chi_{\lambda_b} \right)$ is now second class and the constraints can be imposed strongly. The new brackets appearing at this level are the same ones of the previous level, in the case of the remaining variables. Also we are left with the following constraints only
\begin{eqnarray}
&&H_{D+1}\approx 0, \qquad
H_{4}=\xi^{4}-M \approx 0, \qquad
P_{\mu}^{b}P_{b}^{\mu}-m_{b}^{2}\approx 0, \nonumber \\
&& \chi_{H_{D+1}}\approx 0, \qquad
\chi_{H_4}\approx 0, \qquad
\chi_{P_b}\approx 0.
\end{eqnarray}
Let us consider now the pair $\left(  H_{D+1},\chi_{H_{D+1}
}\right)$. In order to fix the gauge, we must have
\begin{equation}
\left\{  H_{D+1},\chi_{H_{D+1}}\right\}  =C(\xi^{4}, z^4, P_a^{\mu},X_a^{\mu}%
)\neq 0.
\label{CMP}
\end{equation}
Conservation in time of the constraint $\chi_{H_{D+1}}$ produces the following  relation
\begin{eqnarray}
-\tilde{\Lambda}C+\tilde{\lambda}\left\{  \chi_{H_{D+1}
},H_{4}\right\}
+\sum_{b}\tilde{\lambda}_{b}\left\{  \chi_{H_{D+1}
},p_{\mu}^{b}p_{b}^{\mu
}\right\}=0.
\end{eqnarray}
Next  we introduce the invertible coordinate transformation
\begin{equation}
\left( \xi^4, P_b^{\mu}, x^4, X_{b \mu}\right)  \rightarrow\left(  p_a^{\mu},x_a^{\mu},H_{D+1},\chi_{H_{D+1}
}\right),
\end{equation}
with
\begin{eqnarray}
\left\{  H_{D+1},x_a^{\nu}\right\}   &  =\left\{  H_{D+1},p_a^{\nu}\right\}
=\left\{  \chi_{H_{D+1}
},x_a^{\nu}\right\}  =\left\{  \chi_{H_{D+1}
},p_a^{\nu
}\right\}=0.
\end{eqnarray}
The Dirac brackets for the surviving variables, arising after  this gauge fixing, remain the same as the  previous ones.
Imposing these two constraints in the Hamiltonian, and after the  redefinition  ${\hat \lambda}={\tilde \lambda}(\xi^4+M)$ we get
\begin{eqnarray}
&& H =\hat{\lambda}\left(  \left(  \xi^{4}\right)  ^{2}-M^{2}\right)
+\sum_{b}\tilde{\lambda}_{b}\left(  p_{\mu}^{b}p_{b}^{\mu}-m_{b}^{2}\right)
=\hat{\lambda}\left(  \sum_{a}p_{\mu}^{a}p_{a}^{\mu}+\kappa^{2}%
-M^{2}\right)  +\sum_{b}\tilde{\lambda}_{b}\left(  p_{\mu}^{b}p_{b}^{\mu
}-m_{b}^{2}\right), \nonumber \\
&&  =\hat{\lambda}\left(  \sum_{a}p_{\mu}^{a}p_{a}^{\mu}-m_{a}^{2}\right)
+\sum_{b}\tilde{\lambda}_{b}\left(  p_{\mu}^{b}p_{b}^{\mu}-m_{b}^{2}\right),
\nonumber \\
&&  =\sum_{a}\tilde{\lambda}_{a}\left(  p_{\mu}^{a}p_{a}^{\mu}-m_{a}%
^{2}\right),   \quad  a=1,...,N, \quad {\tilde \lambda_1}={\hat \lambda}.
\end{eqnarray}
At this stage, the model is invariant under
$SO(3,1)$.

\subsection{ The  Snyder basis}
To determine this basis we choose the gauge fixing
\begin{equation}
\chi_{H_{D+1}}=z_{4}\xi^{4}+\sum_{a} X_{\mu}^{a}P_{a}^{\mu}-T,
\end{equation}
which produces $C=2\kappa^2$ from Eq. (\ref{CMP}). The new coordinates in phase space are
\begin{eqnarray}
&&p_{a}^{\nu}=\kappa \frac{P_{a}^{\nu}}{\xi^{4}}, \qquad
x_{\mu}^{a}=\frac{1}{\kappa}\left(  X_{\mu}^{a}\xi_{4}-z_{4}p_{\mu}%
^{a}\right),\label{DEFSC}
\end{eqnarray}
having zero brackets with $H_{D+1}$ and $\chi_{H_{D+1}}$. Solving for $ x_{\mu}^{a} ,p_{a}^{\nu}$
results in
\begin{eqnarray}
&&p_{a}^{2}=\kappa^{2}\frac{P_{a}^{2}}{\left(  \xi^{4}\right)^{2}}, \\
&& \sum_{a}p_{a}^{2}=\frac{\kappa^{2}}{\left( \xi^{4} \right)^{2}}
\sum_{a}P_{a}^{2}= \frac{\kappa^2}{\left( \xi^4 \right)^2}\sum_{a}m_{a}^{2}= \kappa^{2}\frac{\kappa^{2}-\left(\xi^{4}\right)^{2}%
}{ \left( \xi^{4} \right)^{2}},\\
&&\xi^{4}=\frac{\kappa}{\sqrt{1+\sum_{a}p_{a}^{2}/\kappa^{2}}},   \\
&& P_{a}^{\nu}=\frac{\xi^{4}p_{a}^{\nu}}{\kappa}=\frac{p_{a}^{\nu}}%
{\sqrt{1+\sum_{c}p_{c}^{2}/\kappa^{2}}},
\end{eqnarray}%
\begin{eqnarray}
&&\sum_{a}x_{\mu}^{a}p_{a}^{\mu}=-X_{\mu}^{a}P_{a}^{\mu}-\frac{z_{4}}%
{\xi^{4}}P_{\mu}^{a}P_{a}^{\mu}=-T+\left(  \left(  \xi^{4}\right)  ^{2}%
-\sum_{a}m_{a}^{2}\right)  \frac{z_{4}}{\xi^{4}}=-T+\frac{\kappa^{2}}{\xi^{4}%
}z_{4},\\
&&z_{4}=\frac{\xi^{4}}{\kappa^{2}}\left(  \sum_{a}x_{\mu}^{a}p_{a}^{\mu
}+T\right) =\frac{\sum_{a}x_{\mu}^{a}p_{a}^{\mu}+T}{\sqrt{\kappa^{2}+\sum
_{c}p_{c}^{2}}},\\
&&X_{\mu}^{a}=\left(  \sqrt{1+\sum_{a}p_{a}^{2}/\kappa^{2}}\, x_{\mu}%
^{a}+\frac{1}{\kappa}\frac{\sum_{c}x_{\nu}^{c}p_{c}^{\nu}+T}{\sqrt{1+\sum
_{a}p_{a}^{2}/\kappa^{2}}}\,p_{\mu}^{a}\right).
\end{eqnarray}
The brackets are calculated from the definitions (\ref{DEFSC}) yielding
\begin{eqnarray}
&&\left\{  x_{\mu}^{a},x_{\nu}^{c}\right\}= -\frac{1}{\kappa^{2}}\left(  x_{\mu}^{a}p_{\nu}^{c}-p_{\mu}%
^{a}x_{\nu}^{c}\right), \quad  \left\{  p_{\mu}^{a},p_{\nu}^{c}\right\}=0, \quad
\left\{  x_{\mu}^{a},p_{\nu}^{c}\right\}=\eta_{\mu\nu}\delta^{ac}-\frac{p_{\mu}^{a}p_{\nu}^{c}%
}{\kappa^{2}}.\label {SNYDERB}
\end{eqnarray}
Let us observe that they reproduce the corresponding brackets (\ref{BRACKETSC}) for the one particle case.

In terms of the new variables, the Hamiltonian is, after some  redefinitions of the arbitrary functions $\tilde \lambda_a$ (which we denote by the same letter),
\begin{eqnarray}
&&H=\sum_{a}\tilde{\lambda}_{a}\left(  P_{\mu}^{a}P_{a}^{\mu}-m_{a}%
^{2}\right)  =\sum_{a}\tilde{\lambda}_{a}\left(  \frac{\kappa^{2}p_{\mu}%
^{a}p_{a}^{\mu}}{\kappa^{2}-\sum_{c}p_{c}^{2}}-m_{a}^{2}\right),\nonumber  \\
&& H =\sum_{a}\tilde{\lambda}_{a}\left(  p_{\mu}^{a}p_{a}^{\mu}-m_{a}^{2}\left(
1-\sum_{c}p_{c}^{2}/\kappa^{2}\right)  \right).
\end{eqnarray}%
We also use the following relations
\begin{eqnarray}
&&\sum_{a}p_{a}^{2}/\kappa^{2}=\frac{1}{(\xi^4)^2}\sum_a P^2_a=\left(\sum_{a}p_{a}^{2}/\kappa^{2}+1\right)  \sum
_{c}m_{c}^{2}/\kappa^{2},\nonumber \\
&&\left(  1-\sum_{c}m_{c}^{2}/\kappa^{2}\right)  \sum_{a}p_{a}^{2}/\kappa^{2}
=\sum_{c}m_{c}^{2}/\kappa^{2},\nonumber \\
&&\sum_{a}p_{a}^{2}=\kappa^{2}\frac{\sum_{c}m_{c}^{2}}{\kappa^{2}-\sum_{c}m_{c}%
^{2}}.
\end{eqnarray}
The final expression is
\begin{equation}
H=\sum_{a}\tilde{\lambda}_{a}\left(  p_{\mu}^{a}p_{a}^{\mu}-m_{a}^{2}\left(
\frac{\kappa^{2}}{\kappa^{2}-\sum_{c}m_{c}^{2}}\right)\right).
\end{equation}
\subsection{ Symmetries }
Since the constraint that defines the Snyder basis is Lorentz invariant  in four dimensions we next verify that the resulting theory   respects such symmetry. We start from the original four-dimensional Lorentz  generators for each particle $L_{\mu\nu}^{a}  =X_{\mu}^{a}P_{\nu}^{a}-X_{\nu}^{a}P_{\mu}^{a}$, which maintain the same form when written in terms of the physical variables
\begin{equation}
L_{\mu\nu}^{a}=x_{\mu}^{a}p_{\nu}^{a}-x_{\nu}^{a}p_{\mu}^{a}.
\end{equation}
Using the brackets (\ref{SNYDERB}) we can calculate the corresponding algebra  which yields
\begin{eqnarray}
 \left\{L_{\mu\nu}^{a},L_{\sigma\tau}^{b}\right\}&=&\delta^{ab}\left(  \eta_{\sigma\nu}\left(  x_{\mu}^{a}p_{\tau}^{b}%
-x_{\tau}^{a}p_{\mu}^{b}\right)  +\eta_{\tau\mu}\left(  x_{\nu}^{a}p_{\sigma
}^{b}-x_{\sigma}^{a}p_{\nu}^{b}\right) \right.\\
&& \left.  +\eta_{\nu\tau}\left(  x_{\sigma}%
^{a}p_{\mu}^{b}-x_{\mu}^{a}p_{\sigma}^{b}\right)  +\eta_{\mu\sigma}\left(
x_{\tau}^{a}p_{\nu}^{b}-x_{\nu}^{a}p_{\tau}^{b}\right)  \right), \nonumber\\
\left\{L_{\mu\nu}^{a},p_{\sigma}^{b}\right\}&=& \delta^{ab}\left( -\eta_{\mu\sigma}p_{\nu}^{a}+\eta_{\nu\sigma}p_{\mu
}^{a}\right), \nonumber \\
\left\{  p_{\mu}^{a},p_{\nu}^{b}\right\} &=&0.
\end{eqnarray}%
Next we sum over  $a$ and  $b$, obtaining
\begin{eqnarray}
\left\{  L_{\mu\nu},L_{\sigma\tau}\right\}&=&\eta_{\sigma\nu}L_{\mu\tau
}-\eta_{\mu\tau}L_{\sigma\nu}-\eta_{\tau\nu}L_{\mu\sigma}+\eta_{\mu\sigma
}L_{\tau\nu}, \\
\left\{  L_{\mu\nu},p_{\sigma}\right\}&=&\eta_{\nu\sigma}P_{\mu}-\eta
_{\mu\sigma}P_{\nu},\qquad
\left\{ p_{\mu},p_{\nu}\right\}=0,
\end{eqnarray}
where $L_{\mu\nu}=\sum_{a} L_{\mu\nu}^{a}, \,\, p_\mu=\sum_{a}p_{\mu}^{a}$.
This  is precisely the  Poincar\'{e} algebra.

\subsection{ The equations of motion }
They are
\begin{eqnarray}
&&\dot{x}_{\mu}^{a}=\left\{x_{\mu}^{a},H\right\}=
2 \left(
-\tilde{\lambda}_{a}+\sum_{b}\tilde{\lambda}_{b}\frac{p_{b}^{\nu}p_{\nu}^{b}%
}{\kappa^{2}}\right)  p_{\mu}^{a}.
\end{eqnarray}%
Again, using
\begin{eqnarray}
&&p_{a}^{2}=\kappa^{2}\frac{P_{a}^{2}}{\left(  \xi^{4}\right)  ^{2}}%
=\kappa^{2}\frac{m_{a}^{2}}{\left(  \xi^{4}\right)^{2}}, \qquad
\left(\xi^{4}\right)^{2}=\kappa^{2}-\sum_{a}m_{a}^{2}, \qquad
p_{a}^{2}
=\frac{\kappa^{2}m_{a}^{2}}{\kappa^{2}-\sum_{c}m_{c}^{2}}%
\end{eqnarray}
we finally obtain
\begin{equation}
\dot{x}_{\mu}^{a}=2\left(  -\tilde{\lambda}_{a}+\frac{\sum_{b}\tilde{\lambda
}_{b}m_{b}^{2}}{\kappa^{2} + \sum_{c}m_{c}^{2}}\right) \, p_{\mu}^{a}.
\end{equation}
The remaining equations are
\begin{eqnarray}
\dot{p}_{\mu}^{a}  &  =\left\{  p_{\mu}^{a},H\right\}=0, \quad \rightarrow \quad {\ddot x}{\,\,}^a_\mu=0.
\end{eqnarray}
To obtain the velocity $v_{i}^{b}$ as a function of time, we have to eliminate the arbitrary functions $\tilde \lambda_a $ which is simply done by defining
\begin{eqnarray}
v_{i}^{b}={P_{i}^{b}}/{P_{0}^{b}}.
\end{eqnarray}
The expressions for the energy and momenta of each particle, which here we assume to be additive, written  in terms of the velocities  are
\begin{eqnarray}
&&P_{a}^{0}=\frac{\kappa m_{a}}{\sqrt{\kappa^{2}+\sum_{c}m_{c}^{2}}}%
\frac{1}{\sqrt{1-\mathbf{v}^{2}}}, \qquad
P_{i}^{a}=\frac{\kappa m_{a}}{\sqrt{\kappa^{2}+\sum_{c}m_{c}^{2}}}.
\frac{v_{i}}{\sqrt{1-\mathbf{v}^{2}}}.
\end{eqnarray}
The rest mass of each particle, that is its energy for $v=0$, is
\begin{equation}
m{_{(a)}}_0 = \frac{m_{a}}{\sqrt{1+\sum_{a}m_{a}^{2}/\kappa^{2}}}.
\end{equation}
The total energy, when a bunch of particles moves with the same velocity, can be used to define the rest mass of the corresponding system, obtaining
\begin{equation}
M_0=\frac{\sum_{a}m_{a}}{\sqrt{1+\sum_{c}m_{c}^{2}/\kappa^{2}}}.
\end{equation}
Let us consider  $N$ identical particles ($m_a=m$). In this case we have the following two limits
\begin{eqnarray}
&& (1) \qquad Nm^{2}\gg\kappa^{2}: \qquad  M_0=\frac{Nm}{\sqrt{1+Nm^{2}/\kappa^{2}}}\simeq\sqrt{N}\kappa, \qquad \\
&& (2) \qquad Nm^{2}\ll \kappa^{2}: \qquad  M_0 \simeq Nm\left(  1-\frac{Nm^{2}
}{2\kappa^{2}}\right).
\end{eqnarray}
The above is an important result which means that, in this model, point like particles have masses  $m_a\leq \kappa$, while composite systems ($N$ particles)  admit masses $M_0 \geq \kappa$.

\section*{Acknowledgements}

JML has been supported by a postgraduate fellowship of Fundacion YPF, Argentina.
LFU is partially supported by projects DGAPA-UNAM: IN111210 and IN109013.
He also acknowledges the hospitality at Facultad de F\'\i sica, PUC, together with a sabbatical fellowship from DGAPA-UNAM. RM acknowledges support from CONICET-Argentina.

\end{document}